\documentstyle[12pt,epsfig]{article}
\def\3{\ss}                                                                                        
                                                                           
\renewcommand{\author}{ }                                                                          
\begin{document}
\vspace{1 cm}
%
%---------- Abbreviations -------------
%
\newcommand{\bec}       {\begin{center}}
\newcommand{\eec}       {\end{center}}
\newcommand{\jpsi}     {\mbox{ J/$\psi$ }}
\newcommand{\jpsitoee} {\mbox{J/$\psi \rightarrow e^{+}e^{-} $}}
\newcommand{\jpsitomm} {\mbox{J/$\psi \rightarrow \mu^{+}\mu^{-} $}}
\newcommand{\ppsi}     {\mbox{$\psi (3685)$}}
\newcommand{\gev}      {\mbox{ ${\rm GeV }$ }}
\newcommand{\pip}      {\mbox{${\rm \pi^{+}}$}}
\newcommand{\pim}      {\mbox{${\rm \pi^{-}}$}}
\newcommand{\ee}       {\mbox{${\rm e^{+}e^{-}}$}}
\newcommand{\mm}       {\mbox{${\rm \mu^{+}\mu^{-}}$}}
\newcommand{\mee}      {\mbox{${\rm m_{ee}}$}}
\newcommand{\mmm}      {\mbox{${\rm m_{\mu\mu}}$}}
\newcommand{\wgp}      {\mbox{${\rm W}$~}}
\newcommand{\qsquare}  {\mbox{${\rm Q^{2}}$}}
\newcommand{\ptsq}     {\mbox{${\rm p_{T}^{2}}$}}
\newcommand{\micron}   {\mbox{${\rm \mu m}$}}
\newcommand{\ccbar}    {\mbox{${\rm c\bar c}$}}
\newcommand{\Acce}     {\mbox{$\italic{A}$}}
\newcommand{\Lumi}     {\mbox{$\italic{L}$}}
\newcommand{\BR}       {\mbox{$\italic{B}$}}
\def\ltap{\raisebox{-.4ex}{\rlap{$\sim$}} \raisebox{.4ex}{$<$}}
\def\gtap{\raisebox{-.4ex}{\rlap{$\sim$}} \raisebox{.4ex}{$>$}}               
%
%
% ---- commands from paul -----
%
\def\ctr#1{{\it #1}\\\vspace{10pt}}
\def\si{{\rm si}}
\def\Si{{\rm Si}}
\def\Ci{{\rm Ci}}
\def\px{p_{_{x}}}
\def\py{p_{_{y}}}
\def\pz{p_{_{z}}}
\def\yjb{y_{_{JB}}}
\def\xjb{x_{_{JB}}}
\def\qjb{\qsq_{_{JB}}}
\def\gap{\hspace{0.5cm}}
\renewcommand{\thefootnote}{\arabic{footnote}}
%
% --------  Title, Date and Authors
%
\begin{flushright}{\large DESY-97-147} \end{flushright}
\vspace{1cm}
\bec{\Large\bf
    Measurement of Inelastic J/$\psi$ Photoproduction at HERA
     }\eec
\vspace{1cm}
\vspace{1.0cm}
\bec{\bf 
    ZEUS Collaboration
    }\eec
\vspace{3cm}
%
% -------- Abstract -------------------
%
\begin{abstract}
We present a measurement of the inelastic, non diffractive 
J/$\psi$ photoproduction cross section in the reaction 
$e^{+} p \rightarrow e^{+} \mbox{J}/\psi X$ with the ZEUS detector at HERA. 
The J/$\psi$ was identified using both the $\mu^{+}\mu^{-}$ and 
$e^{+}e^{-}$  
decay channels and events were selected within the range $0.4<z<0.9$ 
($0.5<z<0.9$) 
for the muon (electron) decay mode, where $z$ is the fraction of the photon 
energy carried by the J/$\psi$ in the proton rest frame.
The cross section, the $p^2_T$ and the $z$ distributions, after 
having subtracted the
contributions from resolved photon and diffractive proton dissociative
processes, are given for the photon-proton centre of mass energy range
$50<W<180$ GeV; $p^2_T$ is the square of the J/$\psi$ 
transverse momentum with respect to the incoming proton beam direction.   
In the kinematic range $0.4 < z < 0.9$ and 
$p^2_T > 1$ GeV$^2$, NLO calculations of the photon-gluon fusion process 
based on the colour-singlet model are in good agreement with the data. 
The predictions of a specific leading order colour-octet model, as 
formulated to describe the CDF data on J/$\psi$ hadroproduction, are not 
consistent with the data.
\end{abstract}
%
%--------  Reset page counter and goto next page
%
\setcounter{page}{0}
\thispagestyle{empty}
\pagenumbering{Roman}                                                                              
\newpage
%===================================================================                               
%                                                                                                  
%  MEMBER NAME  AUTH49 (ZEUS)     M  TEX                                                           
%                                                                                                  
%  JH.: transformed to a format, which is suited as input for                                      
%       CONVERT, which automatically creates author-indices                                        
%                                                                                                  
%  Don't remove lines starting with a percent sign %,                                              
%  CONVERT may need them urgently !                                                                
%                                                                                                  
%=====================================================================     
\begin{center}                                                                                     
{                      \Large  The ZEUS Collaboration              }                               
\end{center}                                                                                       
  J.~Breitweg,                                                                                     
  M.~Derrick,                                                                                      
  D.~Krakauer,                                                                                     
  S.~Magill,                                                                                       
  D.~Mikunas,                                                                                      
  B.~Musgrave,                                                                                     
  J.~Repond,                                                                                       
  R.~Stanek,                                                                                       
  R.L.~Talaga,                                                                                     
  R.~Yoshida,                                                                                      
  H.~Zhang  \\                                                                                     
 {\it Argonne National Laboratory, Argonne, IL, USA}~$^{p}$
% ...
\newpage                                                                                           
% ...
$^{\    2}$ now at Univ. of Salerno and INFN Napoli, Italy \\                                      
% ...
$^{  23}$ now at ATLAS Collaboration, Univ. of Munich \\                                           
% ...
$^{  33}$ partially supported by the Foundation for German-Russian Collaboration                   
DFG-RFBR \\ \hspace*{3.5mm} (grant nos 436 RUS 113/248/3 and 436 RUS 113/248/2)\\                  
% ...
                                                           %                                       
                                                           %                                       
% \par         % if index listing & table fit to 1 page, put gap here                              
\newpage   % alternatively: go to newpage, if page is too small                                    
                                                           %                                       
% \institute_references_start    % do not touch or move this line !                                
                                                           %                                       
\begin{tabular}[h]{rp{14cm}}                                                                       
$^{a}$ &  supported by the Natural Sciences and Engineering Research                               
          Council of Canada (NSERC)  \\                                                            
$^{b}$ &  supported by the FCAR of Qu\'ebec, Canada  \\                                            
$^{c}$ &  supported by the German Federal Ministry for Education and                               
          Science, Research and Technology (BMBF), under contract                                  
          numbers 057BN19P, 057FR19P, 057HH19P, 057HH29P, 057SI75I \\                              
$^{d}$ &  supported by the MINERVA Gesellschaft f\"ur Forschung GmbH,                              
          the German Israeli Foundation, and the U.S.-Israel Binational                            
          Science Foundation \\                                                                    
$^{e}$ &  supported by the German Israeli Foundation, and                                          
          by the Israel Science Foundation                                                         
  \\                                                                                               
$^{f}$ &  supported by the Italian National Institute for Nuclear Physics                          
          (INFN) \\                                                                                
$^{g}$ &  supported by the Japanese Ministry of Education, Science and                             
          Culture (the Monbusho) and its grants for Scientific Research \\                         
$^{h}$ &  supported by the Korean Ministry of Education and Korea Science                          
          and Engineering Foundation  \\                                                           
$^{i}$ &  supported by the Netherlands Foundation for Research on                                  
          Matter (FOM) \\                                                                          
$^{j}$ &  supported by the Polish State Committee for Scientific                                   
          Research, grant No.~115/E-343/SPUB/P03/002/97, 2P03B10512,                               
          2P03B10612, 2P03B14212, 2P03B10412 \\                                                    
$^{k}$ &  supported by the Polish State Committee for Scientific                                   
          Research (grant No. 2P03B08308) and Foundation for                                       
          Polish-German Collaboration  \\                                                          
$^{l}$ &  partially supported by the German Federal Ministry for                                   
          Education and Science, Research and Technology (BMBF)  \\                                
$^{m}$ &  supported by the Fund for Fundamental Research of Russian Ministry                       
          for Science and Edu\-cation and by the German Federal Ministry for                       
          Education and Science, Research and Technology (BMBF) \\                                 
$^{n}$ &  supported by the Spanish Ministry of Education                                           
          and Science through funds provided by CICYT \\                                           
$^{o}$ &  supported by the Particle Physics and                                                    
          Astronomy Research Council \\                                                            
$^{p}$ &  supported by the US Department of Energy \\                                              
$^{q}$ &  supported by the US National Science Foundation \\                                       
\end{tabular}                                                                                      
                                                           %                                       
% \institute_references_end     % do not touch or move this line !                                                                             
%
%--------  Reset page counter and goto next page
%
%\thispagestyle{empty}
\newpage
\setcounter{page}{1}
\pagenumbering{arabic}                                                                              
%
%--------  Introduction and Motivation
%
\section{\bf Introduction}
\label{s:intro}

\begin{figure}[p]
\epsfig{file=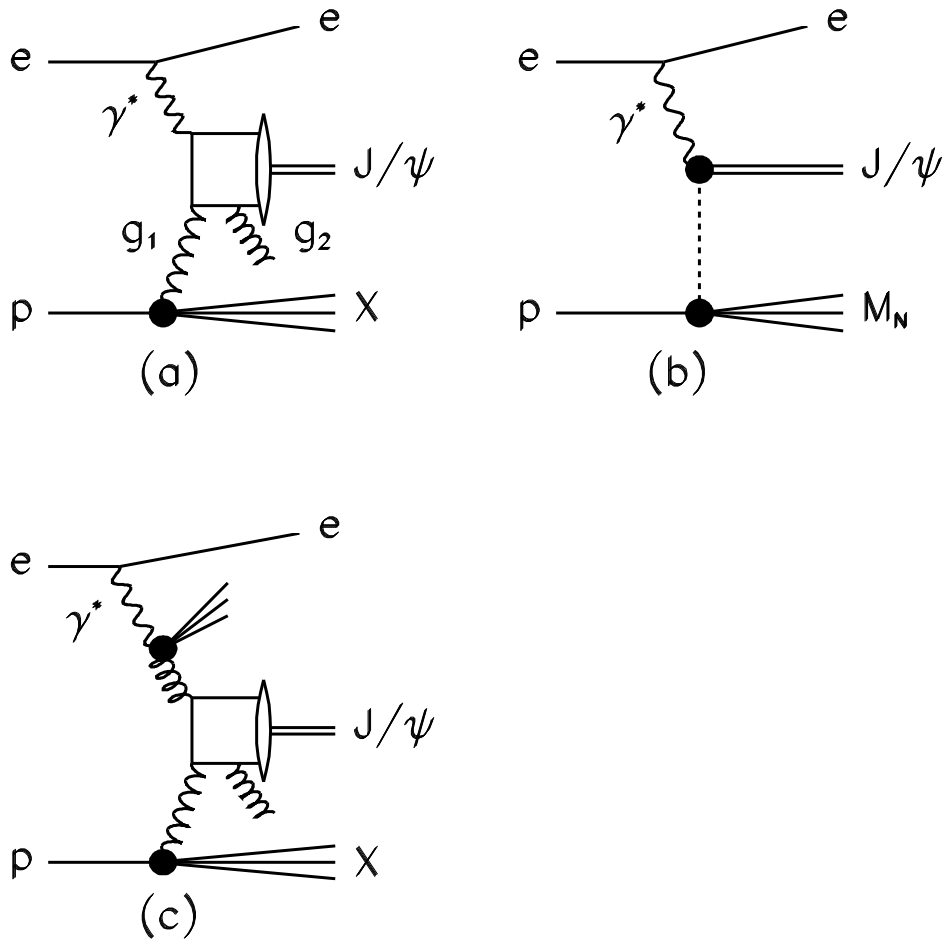,%
      width=15cm%
        }
\caption{Dominant inelastic J/$\psi$ production mechanisms at HERA. 
Photon-gluon fusion is described by diagram (a).
Diagrams (b) and (c) correspond to diffractive proton dissociation and
resolved photon J/$\psi$ production, respectively.}
\label{Fig:Feyn}
\end{figure}

The inelastic reaction $e^{+} p \rightarrow e^{+} \mbox{J}/\psi X$ 
in the photoproduction regime ($Q^2 \approx 0$ GeV$^2$, where $Q^2$ is the
photon virtuality) is thought to proceed via direct photon-gluon 
fusion, diffractive proton dissociation or resolved photon processes. 
These three 
possibilities are shown in Fig.~\ref{Fig:Feyn}. In this paper we are primarily 
interested in the contribution from the direct photon-gluon fusion process, 
shown in Fig. 1a, for which full next-to-leading order (NLO) QCD calculations 
are available \cite{kramer}, in the framework of the colour-singlet 
model \cite{bj}. 
The predicted cross section is sensitive to the gluon density in the 
proton at an energy scale corresponding approximately to the heavy quark mass. 
Perturbative QCD calculations using as input the parton densities extracted
from other processes can therefore provide a consistency test for QCD. 
In inelastic J/$\psi$ photoproduction the concepts of direct 
and resolved photon contributions 
remain distinct up to the NLO level. This is due to the particular 
spin and colour state projection involved in the calculation,  
which makes this reaction different from those involving light 
and open heavy quark production, in which only the sum 
of direct and resolved terms is unambiguously defined at NLO \cite{kramer2}.
 
The different processes in Fig.~\ref{Fig:Feyn} can be distinguished 
by means of the inelasticity 
variable $z$, the fraction of the photon energy carried by the 
J$/\psi$ in the proton rest frame \cite{terron}. The diffractive proton
dissociation process dominates the high $z$ region ($z > 0.9$), the
photon-gluon fusion the intermediate $z$ region,  
while the resolved photon process is dominant in the low $z$ region. 
There are also differences in
the $p_T$ distribution, $p_T$ being the J/$\psi$ transverse 
momentum with respect to the incoming proton beam direction. Compared to 
photon-gluon fusion and resolved photon processes the diffractive reaction    
produces J/$\psi$ mesons with relatively low $p_T$ ($p_T~ \ltap ~1$ GeV).

The direct photon-gluon fusion process is described in 
\cite{kramer} in the framework of the colour-singlet model 
\cite{bj}, in which the initial state photon and gluon interact 
giving a final state $c\bar{c}$ pair with the J/$\psi$ quantum numbers 
through the emission of a hard gluon in the final state 
($\gamma~+~g_1 \rightarrow \mbox{J}/\psi~+~g_2$). When a similar 
model was used to study J$/\psi$ hadroproduction \cite{br} at the Tevatron, 
the predictions, at lowest order in the strong coupling constant $\alpha_s$, 
underestimated the data 
\cite{CDF} by about one order of magnitude. The measured  
cross section could be explained  in part by adding 
colour-octet contributions \cite{octet}. 
In this case the $c\bar{c}$ pair is produced in a colour-octet 
state (short distance process) and later binds to form a 
J/$\psi$ (long distance process). 
While the short distance terms are calculable through perturbative QCD, 
the long distance terms are nonperturbative and have to be determined from 
the data themselves. It is therefore interesting to look for evidence 
of the octet mechanism at HERA, where it is expected to contribute 
at high $z$ \cite{cac}.

Previous fixed target experiments both in the photoproduction \cite{FTPS,NA14} 
and in the electroproduction \cite{BFP,EMC1,NMC,EMC2} regime measured the 
inelastic \jpsi cross section for photon-proton centre of mass energies, $W$, 
between 10 and 20 GeV. 
The H1 Collaboration \cite{H1} has published results on inelastic 
\jpsi production in the interval $30 < W < 150$ GeV.

In the following sections, after a brief description of the experimental 
conditions, we discuss the kinematics of inelastic \jpsi production and 
the criteria used to select events in the region where the direct 
photon-gluon fusion process is dominant. We then 
evaluate the cross section for this process in the range 
$50 < W < 180$ GeV and $0.4 < z < 0.9$ for the muon case and in 
$90 < W < 180$ GeV, $0.5 < z < 0.9$ for the electron case. 
The cross section is extrapolated 
to $z = 0$ assuming the direct photon-gluon fusion model. 
Comparisons with NLO calculations are discussed in section \ref{s:results} for 
the restricted kinematic range 
$z<0.8$ and $p^2_T > 1$ GeV$^2$, where the calculations are reliable. 
 
The data were collected in 1994 and correspond
to an integrated luminosity of 2.99$\pm$0.05 pb$^{-1}$.

\section{Experimental Conditions}
\label{Sect:ExptCond}

\subsection{\bf HERA}
\label{s:hera}

During 1994 HERA operated with a proton beam energy of 820~GeV and a
positron beam energy of 27.5~GeV.
There were 153 colliding proton and positron bunches 
together with an additional 17 unpaired proton bunches and 15 unpaired
positron bunches.
The root mean square (rms) proton bunch length was approximately 20~cm
while the positron bunch length was small in comparison.
The time between bunch crossings was 96~ns.
The typical instantaneous luminosity was 
$1.5 \times 10^{30}$~cm$^{-2}$~s$^{-1}$. 

\subsection{\bf The ZEUS Detector}
\label{s:zeus}

The main ZEUS detector components used in this analysis are outlined below.
A detailed description of the ZEUS detector can be found 
elsewhere \cite{BlueBook}.
In the following the ZEUS coordinate system is used, the $Z$ axis of 
which is coincident with the nominal proton beam axis, the $X$ axis is 
horizontal and points towards the centre of HERA and the $Y$ axis
completes a right-handed coordinate system. 
The origin of the coordinates is at the nominal interaction point.

The momenta and trajectories of charged particles are reconstructed using the
vertex detector  (VXD) \cite{VXD} and the central tracking
detector (CTD)~\cite{CTD}. 
The VXD and the CTD are cylindrical drift chambers covering the angular
region $15^o < \theta < 164^o$ (where $\theta$ is the polar angle with respect
to the proton direction).
The chambers are located in a magnetic field of 1.43 T produced by a thin 
superconducting solenoid. 

The high resolution uranium-scintillator calorimeter
(CAL)~\cite{CAL} surrounding the coil is 
divided into three parts, the forward calorimeter (FCAL), the barrel 
calorimeter (BCAL) and the rear calorimeter (RCAL). They cover polar angles 
from  $2.6^o$ to $36.7^o$, $36.7^o$ to $129.1^o$, and $129.1^o$ to $176.2^o$, 
respectively. 
Each part consists of towers which are longitudinally subdivided into 
electromagnetic (EMC) and hadronic (HAC) readout cells. The CAL also provides 
a time resolution of better than 1 ns for energy deposits greater than 4.5 GeV,
and this timing is used for background rejection.

The hadron electron separator (HES) \cite{HES} consists of silicon detectors 
400 $\mu\mbox{m}$ thick. In the 1994 running period only the rear part (RHES) was 
operational. The RHES is located in the RCAL at a depth of 3.3 radiation 
lengths, covering an area of about 10 m$^2$. Each silicon pad has an area 
of 28.9 x 30.5 mm$^2$, providing a spatial resolution of about 9 mm for a
single hit pad. If more than one adjacent pad is hit by a shower, a 
cluster consisting 
of  at most 3 x 3 pads around the most energetic pad is considered. 
This allows a more precise reconstruction of the position with a 
resolution of about 5 mm for energies greater than 5 GeV. The RHES measures 
the energy deposited by charged particles near the maximum of an 
electromagnetic shower. 

The muon detectors \cite{MUON}, situated outside the calorimeter, 
consist of limited streamer tube (LST) planes 
with the inner chambers in front of the magnetised iron yoke 
and the outer chambers behind it. 
Owing to the low momentum of the \jpsi decay muons, only the inner chambers 
(BMUI and RMUI) were used in the present analysis. 
The BMUI and the RMUI cover the polar angular 
ranges $34^o < \theta < 135^o$ and $134^o< \theta < 171^o$, respectively.

The luminosity is determined from the rate of events due to the 
Bethe-Heitler process $e^{+} p \rightarrow e^{+} \gamma p$, where the photon 
is measured by the calorimeter of the luminosity detector (LUMI) 
located in the HERA tunnel in the 
direction of the outgoing positron beam \cite{LumiCalc}. 
For the measurements 
presented in this paper the luminosity was determined with a precision of 1.5\%.

\section{Kinematics}
\label{s:kine}

Schematic diagrams for the reaction:
\begin{equation}
    e^{+}(k)p(P) \rightarrow e^{+}(k') \mbox{J}/\psi(p_{J/\psi})X,
\label{e:reaction}    
\end{equation}
where each symbol in parentheses denotes the four-momentum of the 
corresponding particle, are shown in Fig. \ref{Fig:Feyn}. 

The kinematics of the inclusive scattering of 
unpolarised positrons and protons is 
described by the positron-proton centre of mass energy squared ($s$)
and any two of the following variables:
\begin{itemize}
    \item $Q^2=-q^2=-(k-k')^2$, the negative four-momentum squared of the 
         exchanged photon; 
    \item $y=(q\cdot P)/(k\cdot P)$, the fraction of the positron energy 
         transferred to the hadronic final state in the rest frame of the 
         initial state proton;
    \item $W^2 = (q+P)^2= -Q^2+2y(k\cdot P)+M^2_p \approx ys$, the
         centre of mass energy squared of the photon-proton system,
         where $M_p$ is the proton mass.
\end{itemize}

Restricting our measurement to photoproduction events where 
the outgoing positron is not in the CAL acceptance, the $Q^2$ value
ranges from the kinematic minimum $Q^2_{min} = M^2_e y^2/(1-y) \sim 10^{-10}$ GeV$^2$, 
where $M_e$ is the electron mass, to the value at which the scattered 
positron starts to be observed in the uranium calorimeter, $Q^2_{max} \sim 4$ 
GeV$^2$. The median $Q^2$ is approximately 10$^{-4}$ GeV$^2$. 

The value of $y$ was determined by the Jacquet-Blondel 
formula \cite{JacBlo}:
\begin{equation}
y \simeq y_{JB} = \frac{\sum_i (E_i - p_{Z_i} )}{2 E_e} ,
\end{equation}
where the sum runs over all the calorimeter cells and $E_i$ is the cell energy, 
$p_{Z_i}$ is equal to $E_i \cos{\theta_i}$, where $\theta_i$ is the polar angle 
of the cell measured with respect to the nominal vertex, and $E_e$ is the 
incoming positron energy.  
The value of $W$ is determined from the relation $W^2 = y_{JB} s$.

For reaction (\ref{e:reaction}) the inelasticity variable, $z$, is defined by
\begin{equation}      
      z   =   \frac{P \cdot p_{J/\psi} }{P \cdot q} \simeq 
      \frac{(E _{J/\psi}- p_{Z_{J/\psi}})}{2 y E_e},
\end{equation}
where $E_{J/\psi}$ is the J/$\psi$ energy and $p_{Z_{J/\psi}}$ is its momentum 
component along the $Z$ direction.  
In the proton rest frame, 
$z$ is equal to $\frac{E_{J/\psi}}{E_{\gamma}}$, where 
$E_{\gamma}$ is the photon energy.
Experimentally $z$ is estimated from:
\begin{equation}
z = \frac{(E_{J/\psi} - p_{Z_{J/\psi}})}{2 y_{JB} E_e} = 
\frac{(E_{J/\psi} - p_{Z_{J/\psi}})}{\sum_i (E_i - p_{Z_i} )}.
\end{equation}
In the estimations of
$y_{JB}$ and $z$, using equations (2) and (4), 
the contribution of the two leptons from the J/$\psi$ decay
was accounted for by including in the sum their momenta as measured in 
the central tracking detectors while discarding their calorimetric deposits.
 
% As mentioned in the introduction the different J/$\psi$ 
% production mechanisms lead to different $z$ distributions.
% The region $z < 0.9$ is dominated by photon-gluon fusion and the region 
% $z > 0.9$ by diffraction. At HERA the contribution from
% the resolved photon process is expected to become significant in the low $z$ 
% region \cite{terron}. 

\section{\bf Event Selection}
\label{s:sele}

The selection of the muon and electron decay channels followed different paths,  
except for the common veto requirements at the first level trigger (FLT), which  
reject proton-gas background events occuring upstream of the nominal 
interaction point and which are therefore out of time with respect 
to the $e^{+} p$ interactions.

\subsection{Muon Mode}
\label{s:musele}

The candidates for the J/$\psi\rightarrow\mu^+\mu^-$ channel were
selected using the three level ZEUS trigger system.
At the FLT a coincidence between track segments
in the CTD, energy deposits in the CAL and hits in the BMUI or RMUI 
was used to select muon candidates.
The CAL was divided in $Z-\phi$ regions ($\phi$ being the 
azimuthal angle around the $Z$ axis) associated with the 
corresponding 
zones of the barrel and rear muon chambers. A signal above threshold 
in one of the CAL regions in conjunction with a hit in the 
associated barrel or rear muon chamber defined a CAL-BMUI/RMUI match.
This regional matching was demanded together with the requirement 
of tracks in the CTD pointing to the nominal vertex. 

At the second level trigger (SLT), the total energy in the calorimeter 
($E_{Tot} = \Sigma_i E_i$) and the $Z$ component of the momentum  
($\Sigma_i p_{Z_i} = \Sigma_i E_i \cos \theta_i$) were calculated.
The sums run over all calorimeter cells $i$ with an energy, $E_i$, 
above threshold at a polar angle, $\theta_i$, measured with respect to the 
nominal vertex. 
In order to remove proton-gas interactions, events with the 
ratio ${\Sigma_i p_{Z_i}}/{E_{Tot}}$ greater than 0.96 were rejected. 
Part of the cosmic ray background was rejected at the SLT by using the time 
difference of the energy deposits in the upper and the lower halves of the 
calorimeter.

At the third level trigger (TLT) a muon candidate was selected when a 
track found 
in the CTD matched both a cluster with a calorimeter energy deposit 
consistent with 
the passage of a minimum ionising particle (a m.i.p. cluster
\footnote{A cluster is defined as a group of contiguous
cells in the CAL with energy above a set threshold.}) and a
track in the inner muon chambers.
An event containing a muon candidate in the rear (barrel) region 
was accepted if the (transverse) momentum of the CTD track exceeded 1 GeV.
  
The TLT algorithm was again applied in the offline analysis, 
but now the results of the full event reconstruction were used. 
The tracks corresponding to the two muons from the J/$\psi$ decay had to 
satisfy the following criteria, where the subscript 1 denotes 
the triggering muon and the subscript 2 the other muon and $p$ indicates 
the momentum of a muon and $p_t$ its transverse momentum: 
\begin{itemize}
\item $p_1 > 1$ GeV (rear region); $p_{t1} > 1$ GeV (barrel region);
\item $p_2 > 1$ GeV;
\item $p_{t1} + p_{t2} > 2.8$ GeV;
\item pseudorapidities\footnote{The pseudorapidity
is defined as $\eta=-\ln\tan(\frac{\theta}{2})$.} $|\eta_{1,2}| < 1.75$;
\item the second muon track has to match a m.i.p. cluster in the CAL.
\end{itemize}
Cosmic rays were rejected by requiring that the two muon tracks 
were not collinear: events with  
$\Omega > 174^{o}$ were rejected, where $\Omega$ is the angle between the two 
tracks at the interaction point.   

The final inelastic data sample was defined by requiring an energy deposit 
greater than 1~GeV in a cone of $35^o$ 
around the forward direction (excluding the calorimeter deposits due 
to the muons). Elastically produced J$/\psi$ mesons were thus excluded. 
The data were further restricted to the $W$ interval 50 to 180 GeV 
where the acceptance is above 10\%. The data sample was
divided into the three categories:
\begin{itemize}
\item  events with $z$ in the interval 0.9 to 1;
\item  events with $z$ in the interval 0.4 to 0.9;
\item  events with $z <0.4$.
\end{itemize}

\begin{figure}[p]
\epsfig{file=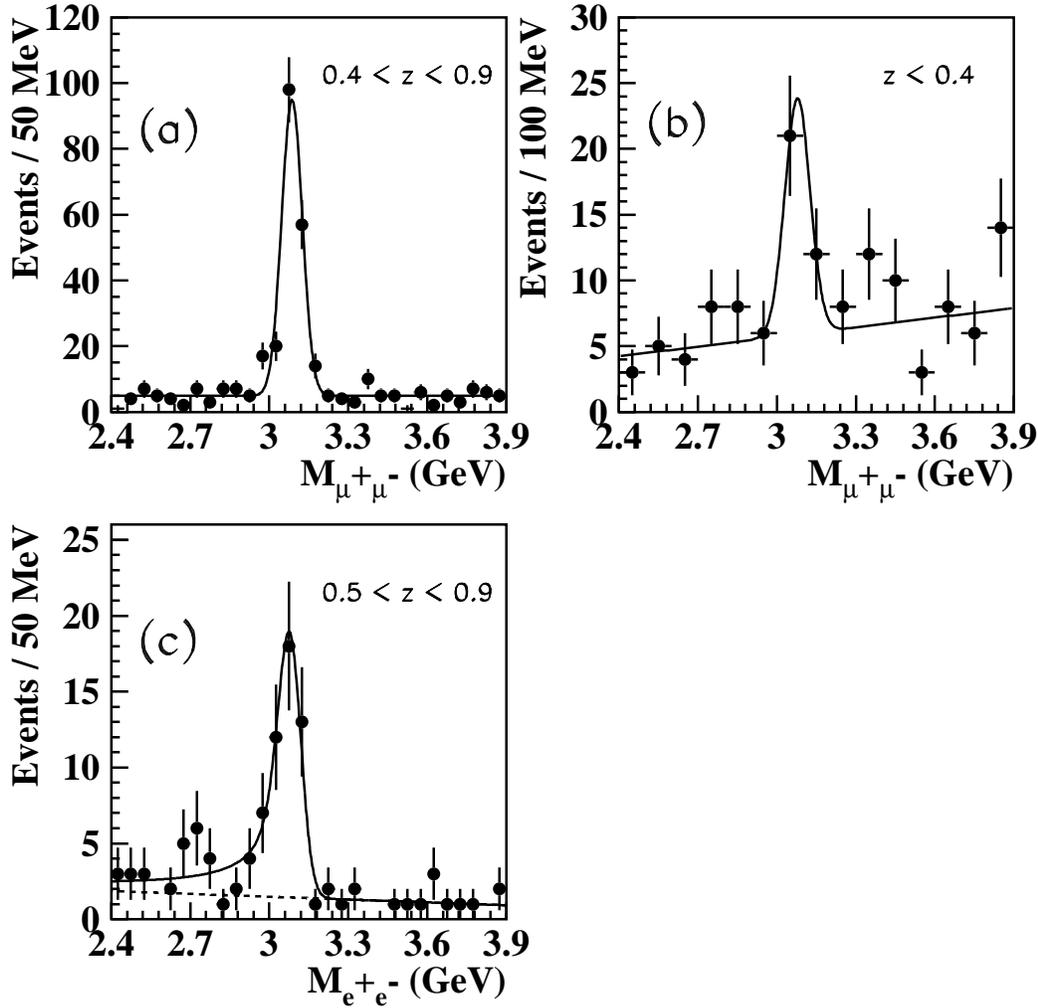,%
      width=15cm%
        }
\caption{Invariant mass spectrum for the muon pair sample (a) for 
$0.4 < z <0.9$ and (b) for $z < 0.4$, in the $W$ range 
50 to 180 GeV. The invariant mass spectrum 
for the electron pair sample ($0.5 < z <0.9$ and $90 < W < 180$ GeV) 
is shown in (c). The muon mass spectrum (a) was fitted to the sum of a 
Gaussian and a flat background; the spectrum (b) was fitted to the sum of a 
Gaussian and a linear 
background. The electron mass spectrum (c) was fitted to the sum of the convolution
of a Gaussian and a bremsstrahlung function plus a linear background.}
\label{Fig:mass}
\end{figure}

The first category is interpreted as coming mainly from the diffractive proton 
dissociation process. The second one is dominated by the photon-gluon 
process (direct process) and the third is a combination of direct and 
resolved processes. The $\mu^{+}\mu^{-}$ invariant mass for the second 
category is shown in Fig. \ref{Fig:mass}a, fitted with a 
Gaussian plus a flat background giving
a mass of 3.086$\pm$0.004 GeV. The rms width is 39$\pm4$ MeV, consistent with the 
Monte Carlo expectations. The invariant mass 
distribution 
for $z<0.4$ events is shown in Fig. \ref{Fig:mass}b. 
Table \ref{Tab:events} contains the fitted number of events above 
background for each category and for various $W$ ranges. 
The data so collected correspond to events with $Q^2 < 4$ GeV$^2$. The
events selected in the chosen $W$ range have $\sum_i(E_i-p_{Z_i}) < 20$ GeV. 
Events with the scattered positron in the CAL are expected to have a 
$\sum_i(E_i-p_{Z_i}) \sim 
2E_e$ = 55 GeV. A cross check with an electron finder confirmed the 
absence of large $Q^2$ events in the sample. 

\subsection{Electron Mode}
\label{s:elesele}

Inelastic \jpsitoee~ candidates were triggered at the FLT 
by demanding the two conditions:

\begin{enumerate}
\item  at least one of the following requirements on the CAL energies:

\begin{itemize}
\item  CAL total energy $>$ 15 GeV;

\item  CAL-EMC energy $>$ 10 GeV;

\item  CAL total transverse energy $>$ 11 GeV;

\item  BCAL-EMC energy $>$ 3.4 GeV;

\item  RCAL-EMC energy $>$ 2.0 GeV;
\end{itemize}

\item  at least one CTD track associated with the nominal vertex. 

\end{enumerate}

At the SLT events were
rejected if $\sum_i p_{Z_i}/E_{Tot}$ was greater than 0.92 
(with $p_{Z_i}$ and $E_{Tot}$ defined as in the previous section). In addition
only events satisfying the following two conditions were accepted:

\begin{itemize}
\item  $\sum_i(E_i-p_{Z_i})$ $>$ 4 GeV, where the sum runs over all calorimeter 
cells $i$;

\item  the sum of the total energy deposits in BCAL-EMC and RCAL-EMC was 
greater than 3 GeV.

\end{itemize}

At the TLT a fast electron identification was carried out by using information 
 from
the CTD and CAL.  Clusters were identified as electrons if
at least 90\% of the cluster's energy was deposited in the electromagnetic 
section.
The tracks from the CTD were extrapolated towards the CAL and matched to the
nearest cluster within 30 cm of the extrapolated track at the CAL face. 
An event was accepted if at least two oppositely 
charged tracks, identified as electrons, were found each with a momentum 
exceeding 0.5 GeV and a transverse momentum 
greater than 0.4 GeV; in addition, the two tracks were required to originate
from points less than 7 cm apart along the $Z$ axis. The 
invariant mass of the track pair, assuming the electron mass for each track, 
had to be greater than 2 GeV. 

The initial offline selection was based on the TLT track-cluster matching 
algorithm, but using the full tracking and CAL information.

Since the transverse momentum ($p_t$) spectrum of the background 
tracks peaks at low $p_t$ values, 
both electron track candidates were required to have
$p_t$ greater than 0.8~GeV. Also, both tracks had to originate 
from the event vertex and satisfy the condition $ | \eta | < 1.75$. 
The large background coming mainly from low energy pions faking 
electrons was further reduced firstly by requiring a tighter matching of 
the tracks to the electromagnetic clusters, with a track-cluster
separation at the CAL face less than 25~cm, secondly by demanding clusters 
with small longitudinal and radial 
dimensions and thirdly by imposing a cut 0.4 $< E_{cluster}/p <$ 1.6, 
where $E_{cluster}$ is the energy of the electromagnetic cluster and $p$ is the 
momentum of the associated track. This 
cut was chosen since, for electrons in the momentum range 1-3 GeV 
(typical for electrons from \jpsi~decays in the present analysis), 
the inactive material in front of the calorimeter means that the 
$E_{cluster}/p$ ratio is about 0.8 with 20\% resolution.

A significant reduction in the remaining background was achieved by using 
the RHES and the information on the specific ionisation energy loss,  
$d{\rm E}/d{\rm x}$, as evaluated from the CTD.
The $d{\rm E}/d{\rm x}$ of a track, calculated from the  
truncated mean of a distribution of pulse amplitudes where 
the lowest 30\% and the highest 10\% were discarded, 
had a resolution of about 12\%, averaged over a broad range in 
$\eta$ ($ | \eta | < 1.5$).  
By requiring $d{\rm E}/d{\rm x}$ for one of the electron candidate 
tracks to be consistent with that expected for an electron,  
93\% of the \ee~ pairs were
retained, while discarding two thirds of the background\footnote{The 
93\% efficiency was computed from $\gamma$ conversions and almost 
background-free \jpsitoee
elastic events.}. Because the identification of electrons via  
$d{\rm E}/d{\rm x}$ is not well understood for low angle tracks, the
use of $d{\rm E}/d{\rm x}$ was limited to tracks with $|\eta|<1$. 
For $\eta <-1$ the RHES can be used for electron identification. 
A HES electron 
cluster was defined as a group of adjacent silicon pads each with an 
energy deposit above 0.6 m.i.p., and the total energy of all pads 
above 5 m.i.p. Track-HES
cluster matching was then performed for tracks already matched to a 
CAL cluster, 
requiring that the distance between the HES cluster and the 
extrapolated track be less than 10 cm. 
The efficiency of the cut on the RHES cluster energy above 5 m.i.p. 
was estimated 
to be 75\% using an almost background-free \jpsitoee elastic sample.

Events were then accepted if the electron tracks satisfied one of the 
following requirements: if both tracks lie in the range $|\eta|<$ 1 they had 
to satisfy the aforementioned $d{\rm E}/d{\rm x}$ cut; if one track was in 
the range $|\eta|<$ 1 and the other in $\eta<-1$ they had to satisfy 
the $d{\rm E}/d{\rm x}$ and the RHES cuts, respectively. All the other 
track combinations were not considered due to the presence of high 
background.   

The final inelastic data sample was defined by requiring an energy deposit 
greater than 1~GeV in a cone of $35^o$ 
around the forward direction (excluding the calorimeter deposits due 
to the electrons). A minimum value of $z$ was required ($z>0.5$) 
to avoid the low $z$ region which is dominated by large 
background. The data were also restricted to the $W$ 
range 90 to 180~GeV, 
where the acceptance is high. 
Using the $z$ variable the electron data sample was
divided in two categories: $z>0.9$ and $0.5 < z < 0.9$.

Figure \ref{Fig:mass}c shows the mass distribution of the electron pairs 
for the second category.
A clear peak at the \jpsi~mass is observed.
The solid line shows an unbinned likelihood fit in which a Gaussian
resolution function has been convoluted with a radiative \jpsi~mass
spectrum and a linear distribution to describe the background (dashed line).
The mass estimated by the fit is $3.089 \pm 0.010$~GeV. 
The rms width is $40 \pm 9$~MeV, consistent with the MC expectation.   
Table \ref{Tab:events} contains, for the two categories and  
$W$ ranges, the fitted number of events above background. 
As for the muon sample, the data so collected correspond to events with 
$Q^2<4$ GeV$^2$. 

\section{\bf Monte Carlo Simulation and Acceptance Calculation}
\label{s:mcacc}

Inelastic J$/\psi$ production from direct photon-gluon fusion 
was simulated using the colour-singlet model 
as implemented in the HERWIG \cite {herw} parton shower generator. 
The range of $Q^2$ was from the kinematic limit ($\approx 10^{-10}$ GeV$^2$) 
to 4 GeV$^2$. 
The energy scale, $\mu^2$, at which the gluon distribution is evaluated 
was chosen to be
$\mu^2=2\hat{s}\hat{t}\hat{u}/(\hat{s}^2+\hat{t}^2+\hat{u}^2)$,
where $\hat{s}$, $\hat{t}$ and $\hat{u}$ are the Mandelstam variables of the 
photon-gluon fusion process.
The mean value of $\mu^2$ is 7 GeV$^2$. 
The gluon structure function in the
proton was parameterized with the MRSD$^{'}$$\_$ \cite{mrs} distribution.
 
For resolved J$/\psi$ production the PYTHIA \cite{pythia} parton shower 
generator was used with the GRV proton \cite{grv} and photon \cite{grvg} 
parton densities. 
The matrix elements for resolved photon processes were computed in the 
colour-singlet framework. 
 
Production of J$/\psi$ mesons accompanied by diffractive proton dissociation
was simulated with EPSOFT \cite{eps}. 
This generator is based on the assumption that the diffractive cross 
section is of the form 
$d\sigma / d|t| dM_N^2\propto {e^{-b_d|t|}}/{M_N^{\beta}}$,  
where $M_N$ is the mass of the dissociative system, and $t$ is the 
four-momentum transfer squared at the proton vertex.  
The value of $b_d$ was chosen to be 1 GeV$^{-2}$ to reproduce the 
observed $p_T^2$ distribution of events with $z>0.9$. For the 
$M_N$ distribution, the value $\beta=2$ was used. 
% The mass of the nucleonic system was generated in the range 
% 1.25 GeV$^2 \leq M_N^2 \leq 0.1~W^2$. 
The simulation  of the dissociative system 
includes a parametrisation of the resonance spectrum.
         
\begin{figure}[p]
\epsfig{file=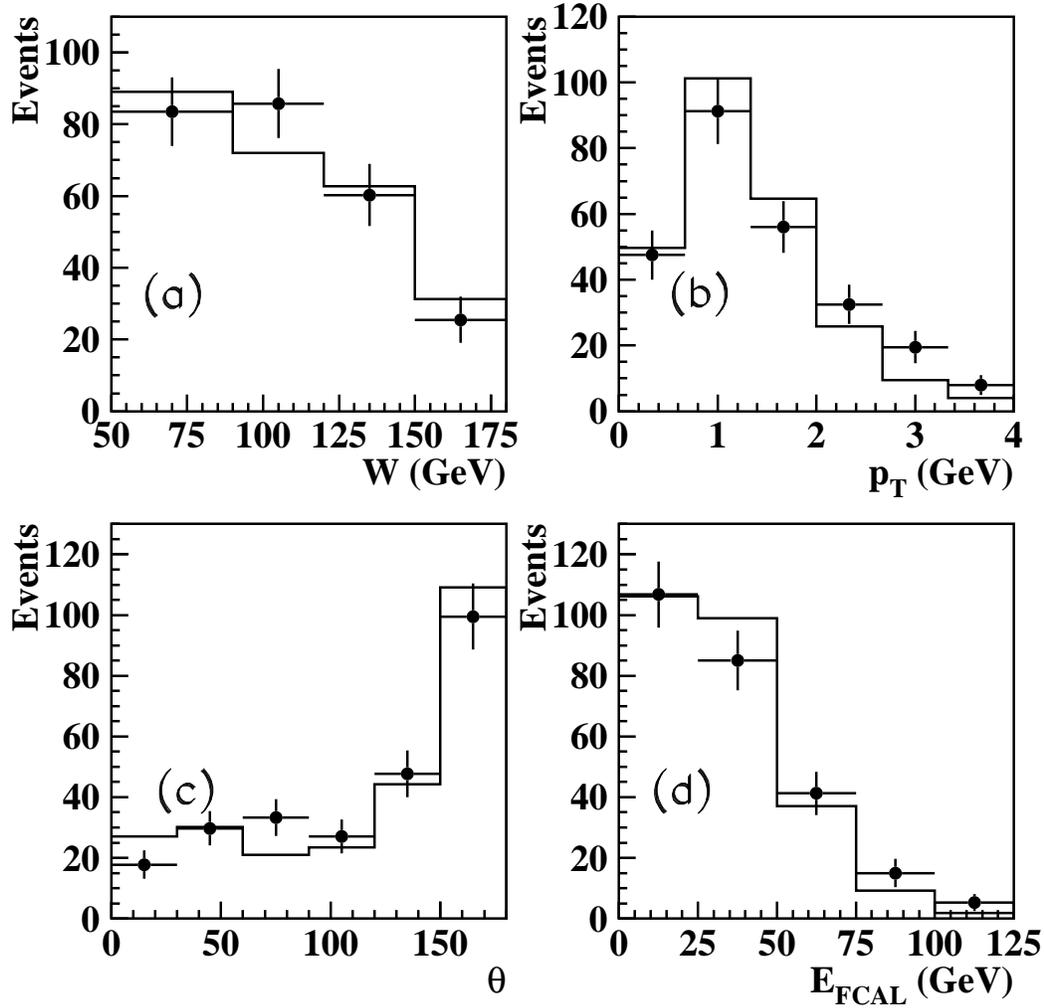,%
      width=15cm%
        }
\caption{In (a) the uncorrected $W$ distribution 
of the $\mu^{+}\mu^{-}$ data (full dots) with $0.4 < z < 1$ is 
compared to the mixture of HERWIG and EPSOFT (continuous histogram) 
described in section \protect\ref{s:mcacc}. 
In (b), (c) and (d) similar comparisons between data and the Monte 
Carlo mixture are shown for the distributions of $p_T$ and 
polar angle $\theta$ of the J/$\psi$ and for the energy in 
the forward calorimeter $E_{FCAL}$, respectively. The Monte Carlo mixture is
normalized to the number of measured events.}
\label{Fig:mix}
\end{figure}

In the muon case a mixture of HERWIG 
$(78^{+4}_{-6})$\% and EPSOFT events gives the best
description of the $z$ distribution with $z$ ranging from 0.4
to 1. For the electron case the percentage is (79$\pm6$)\%.
The same mixture also describes well all the reconstructed kinematic
variables (see Fig. \ref{Fig:mix} for some examples).
The resolution in $z$ is $2\%$ at $z=0.9$ and increases to 
12\% at $z=0.4$. 
The measured values of $z$ suffer from a systematic shift due 
to the energy loss in the inactive material in front of the calorimeter 
and to the undetected particles escaping in the beampipe. This shift is 
20\% at $z=0.4$ and becomes negligible at $z=0.9$. The shift
of $z$ was corrected using the HERWIG Monte Carlo.

The acceptance was estimated as the ratio of the number of accepted Monte Carlo
photon-gluon fusion events to the number generated in the selected 
kinematic range. 
The acceptance, calculated in this manner, accounts for the geometric
acceptance, for the detector, trigger and reconstruction efficiencies, and 
for the detector resolution. Table \ref{Tab:xsect} reports the 
acceptances in various $W$ ranges determined for each decay mode.

\section{Backgrounds to the photon-gluon fusion process}
\label{s:back}
In this section we discuss all the resonant processes which are backgrounds to
the photon-gluon fusion process. 
To calculate integrated and differential cross sections 
the analysis was restricted 
to the region $0.4 < z < 0.9$ and $50 < W < 180$ 
GeV for the muon mode, and to the region $0.5 < z < 0.9$ and  
$90 < W < 180$ GeV for the electron mode. The upper $z$ cut 
is necessary to exclude diffractive J/$\psi$ production; 
the lower cut restricts the data to a range 
with low background and 
where the photon-gluon process is expected to dominate over the
resolved production. 
Table \ref{Tab:events} reports 
the numbers of events coming from the fit to the J/$\psi$ mass peak, 
divided into four $W$ intervals for the muon mode and three intervals 
for the electron mode.  
These numbers include proton diffractive dissociation 
events 
which migrated from the region above $z=0.9$. The background from 
diffractive events was estimated to be 
$f^{diff}= (8\pm2)$\% in the muon mode and 
$f^{diff}= (4\pm1)$\% in 
the electron mode using EPSOFT
and following the method explained in section \ref{s:mcacc}. 
The cross sections were corrected for 
the estimated fraction of proton dissociative events. The difference in the 
size of the contamination between muon and electron decay mode reflects 
the different $W$ range covered. 

At the lower end of the $z$ range one has to consider 
resolved J/$\psi$ photoproduction events for $z > 0.4(0.5)$ as well as 
migration of resolved events 
from $z<0.4~(0.5)$ into the region studied. 
In Fig. \ref{Fig:mass}b the invariant mass plot of the muon decay 
mode for events with $z<0.4$ and $ 50 < W < 180$ GeV is shown. 
The estimated number of
J/$\psi$ events is $19\pm 6$. 
As determined from the $z$ shape of resolved and direct photon Monte Carlo 
generated events, about 50\% of the detected events can be attributed to 
resolved photoproduction. 
Their contribution in the $0.4 < z < 0.9$ range, after diffractive proton 
dissociation subtraction, is $3^{+3}_{-2}$\%. This 
background has been subtracted from the signal events assuming the $W$ 
and $z$ dependence given by PYTHIA Monte Carlo. Due to the small statistics 
of the electron sample the muon result was also used in the electron case. 

Production of $\psi'$ mesons with subsequent decay into J/$\psi$ is a 
contribution not included in the simulation. 
It is estimated in \cite{kramer}, through phase space considerations, to be 
15\% of the J/$\psi$ integrated cross section. 
This result is in good agreement with estimations made by using the value 
of the $\psi'$ to J/$\psi$ ratio coming from low energy data \cite{FTPS,NA14}. 
This contribution was not subtracted. 

\vskip 2.cm
\section{Systematic Errors}
\label{s:sys}
Several factors contribute to the systematic errors in the 
inelastic J/$\psi$ cross section measurement. In the following they are 
divided in two categories: 
{\it decay channel specific errors} and 
{\it common systematic errors}. The first category contains
systematic errors specific to the electron or muon decay channel,
while the second contains systematic errors common to both decay modes.

\paragraph
{\parindent 0pt
{\bf Decay channel specific errors}:
}

\begin{itemize}

\item {\it Trigger}: 
for the muon mode the principal source of uncertainty is the FLT calorimeter 
trigger. The corresponding 
error was estimated by using independent muon triggers, which 
use different calorimeter trigger logic or do not use the calorimeter at all. 
Also for the electron mode the uncertainty is dominated by the FLT calorimeter 
simulation and the corresponding error was estimated by using 
independent calorimeter triggers. 
The size of the error depends on the $W$ range and is of the order 
of $\pm4\%$ in the lowest $W$ bin and of $\pm1\%$ in the highest one. 

\item {\it Event selection}:
this class comprises the systematic errors due to the uncertainties in 
the measurement of momentum, transverse momentum and $\eta$ of the leptonic 
tracks. For the muon channel this class contains also the
uncertainties coming from the $p_{t1} + p_{t2} > 2.8$ GeV cut and the 
collinearity cut. For the electron channel uncertainties in the definition 
of an electron cluster also contribute. Each cut was varied within a range 
determined by the resolution of the variable to which the cut is applied 
and the different contributions obtained were summed in quadrature. 
This error amounts to $\pm2\%$, almost independent of $W$.

\item {\it Muon chamber efficiency}: 
the systematic error ($\pm2\%$) attributed to
uncertainties in the muon chamber reconstruction efficiency was estimated using
cosmic ray events.

\item {\it d$E$/d$x$}: 
the error takes into account the uncertainty of the variation in the 
efficiency of the d$E$/d$x$ cut 
as a function of the track's polar angle, not reproduced by the Monte Carlo
simulation. The size of the error depends on the $W$ range and is of the order
of $\pm7\%$ in the lowest $W$ bin and of $\pm5\%$ in the highest one.

\item {\it RHES}:
the error (+1\%) was estimated by rising 
the cut value from 5 m.i.p. to 6 m.i.p. in the Monte Carlo 
simulation only. This was done to take into account 
possible differences of the calibration in the simulation and in the data. 

\item {\it Fitting procedure}:
different fitting procedures for the J/$\psi$ mass peak were applied 
to the electron channel and the results were the same within the 
statistical uncertainty. The size of the error is --3\%.

\item {\it Branching Ratio}:
the error on the branching ratio J/$\psi \rightarrow l^{+}l^{-}$ is used 
as quoted in \cite{PDG} ($\pm3.2\%$).
\end{itemize}

{\parindent 0pt
{\bf Common systematic errors}:
}

\begin{itemize}
\item {\it Parton density}:
the uncertainty in the acceptance resulting from the uncertainty 
in the form of the gluon distribution was
estimated by changing the default gluon distribution (MRSD$^{'}$$\_$ \cite{mrs}) 
with others (GRV \cite{grv}, MRSG \cite{mrs2}, MRSA$^{'}$ \cite{mrs2}). This 
error is of the order of $\pm1\%$ in all the $W$ ranges, except in the highest 
$W$ bin where it contributes --8\%.

\item {\it Energy scale}: 
the calorimeter energy scale in the Monte Carlo was varied by $\pm 5$\%. 
This affects mainly the $W$ and $z$ determinations and gives a $\pm2\%$ 
contribution. 

\item {\it Proton dissociation}:
The parameters $\beta$ and $b_d$ were varied from $\beta=2$ to $\beta=2.5$
($\beta=2.20\pm0.03$ is the result of \cite{CDFpomeron}) and from $b_d=0.9$ 
GeV$^{-2}$ to $b_d=1.3$ GeV$^{-2}$ (from the analysis of the $p_T^2$ 
distribution of the events with $z > 0.9$). 
This systematic uncertainty is dominated by the 
number of events with $z>0.9$ and is concentrated in the range $50<W<90$ GeV, 
where it gives a contribution of $\pm5\%$, while in the other bins 
it is of the order of the $\pm 1\%$ or lower.

\item {\it Resolved photon}:
this systematic error contains contributions from the limited statistics 
of the muon sample for $z < 0.4$ and the uncertainties in the 
Monte Carlo modelling of the resolved process. The corresponding error 
was evaluated to be $\pm4\%$, independent from $W$.

\item {\it $z$ extrapolation}:
the cross sections were measured down to $z=0.4$ for the muon channel and 
to $z=0.5$ for the electron channel and then extrapolated to zero using the  
HERWIG Monte Carlo. The uncertainty ($\pm3.5\%$) on the extrapolation 
was evaluated by varying $\Lambda_{QCD}$ and the charm mass, $m_c$, 
in the NLO calculation.

\item {\it Angular distribution}:
The angular distribution of the decay 
leptons was modelled using the form ($1+ \alpha \cos^2{\theta^{*}}$), where
$\theta^{*}$ is the decay angle of the leptons in the J/$\psi$ rest frame with
respect to the direction of the J/$\psi$ momentum in the laboratory. 
The data are best described with $\alpha=0$, that is 
a flat distribution. As a systematic check $\alpha$ was varied by one 
standard deviation (i.e. up to $\alpha=0.5$); this gave an error growing from
+5\% in the lowest $W$ bin to +8\% in the highest $W$ bin.  

\item {\it Luminosity}:
as indicated in section \ref{s:zeus}, the uncertainty on the luminosity 
determination is $\pm$1.5\%.

\end{itemize}

The total systematic error, given by the sum in 
quadrature of all the common and uncommon systematic errors, is of the order 
of $\pm10$\% ($\pm15$\%) for the muon (electron) decay channel.

\section{Results}
\label{s:results}

\subsection{Cross Section Calculation}
\label{s:section}

The electroproduction cross section for inelastic J/$\psi$ production,  
after subtracting the contributions of diffractive proton dissociation and 
of resolved photon processes, is calculated as:
\begin{equation}
\sigma(e^{+} p \rightarrow e^{+} \mbox{J}/\psi X) = \frac{N_{evt}}
{\cal{A} ~\cal{L} ~\cal{B}} ,
\end{equation}
where $N_{evt}$ denotes the background subtracted number of J/$\psi$ 
signal events, $\cal{A}$ the acceptance, 
$\cal{L}$ the integrated luminosity and $\cal{B}$ the J/$\psi$ 
leptonic branching fraction 
\cite{PDG}, namely $(6.01\pm0.19)$\% for $\mu^{+}\mu^{-}$ and 
$(6.02\pm0.19)$\% for $e^{+}e^{-}$. 
The photoproduction cross section is related to the $ep$ cross section by
\cite{Flux}
\begin{equation}
    \sigma_{\gamma p \rightarrow J/\psi X} =
     \frac{\int^{y_{max}}_{y_{min}} 
           \int^{Q^2_{max}}_{Q^2_{min}(y)} \Phi(y, Q^2) 
           \sigma_{\gamma p \rightarrow J/\psi X}(y, Q^2) dy dQ^2}
          {\Phi_T}            =
     \frac{\sigma_{e^{+} p \rightarrow e^{+} J/\psi X}}{\Phi_T} ,
\end{equation}
where $\sigma_{\gamma p \rightarrow J/\psi X}$ is 
the mean cross section
in the measured range of $W$ (corresponding to the 
limits $y_{min}$, $y_{max}$) and $Q^2$.
The effective flux, $\Phi_T$, of virtual photons 
from the positron is computed as:
\begin{equation}
\Phi_T = \int^{y_{max}}_{y_{min}}\int^{Q^2_{max}}_{Q^2_{min}(y)}\Phi(y, Q^2) 
dy dQ^2  = \int^{y_{max}}_{y_{min}}\int^{Q^2_{max}}_{Q^2_{min}(y)} 
\frac{\alpha}{2 \pi y Q^2} [1 +(1-y)^2 - \frac{2M_e^2 y^2}{Q^2}] dy dQ^2,
\end{equation}
where $\alpha$ is the electromagnetic coupling constant.
The integrals run from $Q^2_{min} = M_e^2 \frac{y^2}{1-y}$ to 
$Q^2_{max}=4$ GeV$^2$ and from $y_{min}=W^2_{min}/s$
to $y_{max}=W^2_{max}/s$ where $W_{min}$ and $W_{max}$ are the minimum and
maximum values of $W$, respectively, in each chosen interval.

\begin{figure}[b]
\epsfig{file=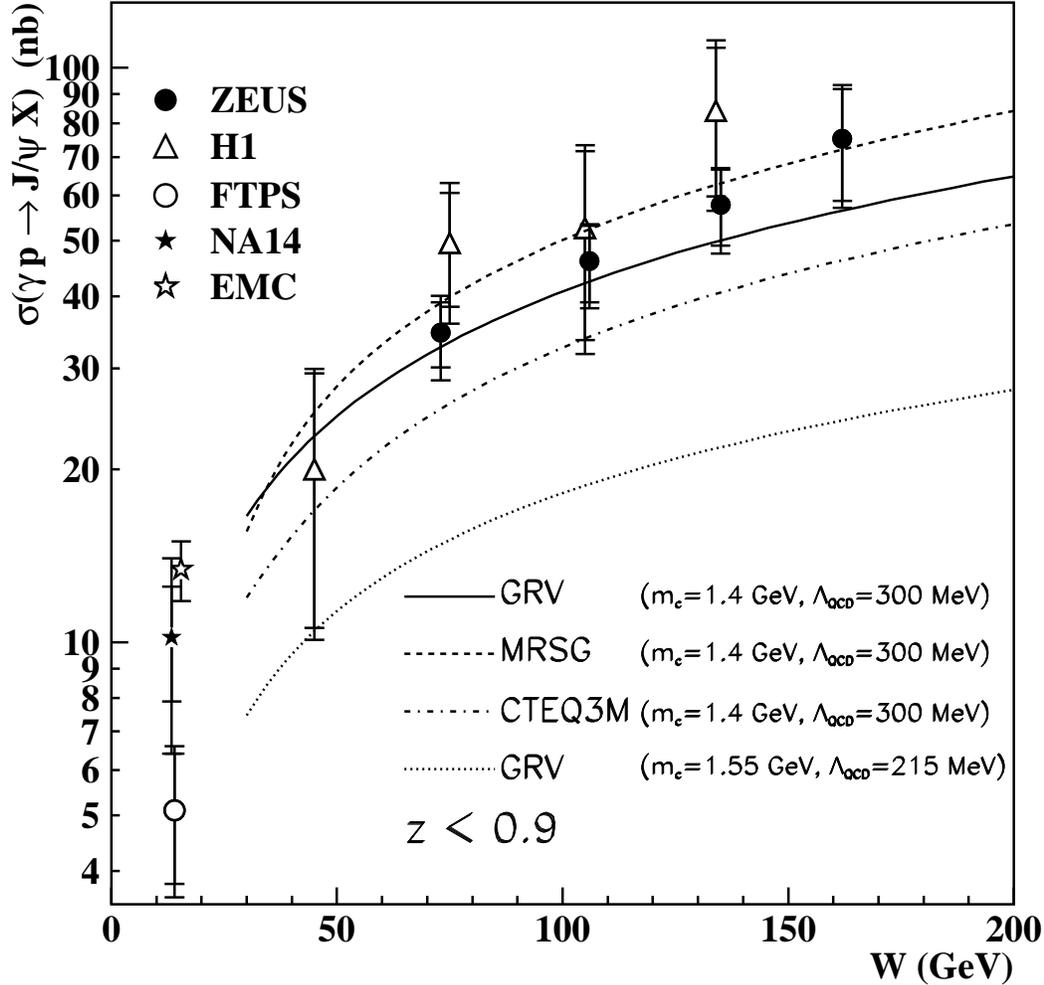,%
      width=15cm%
        }
\caption{The direct inelastic J/$\psi$ photoproduction cross section as a 
function of $W$ for $z<0.9$. 
Data from ZEUS, H1 \protect\cite{H1}, FTPS  
\protect\cite{FTPS}, NA14 \protect\cite{NA14} and EMC \protect\cite{EMC2} 
are shown. The ZEUS result at the lowest $W$ value is obtained with the 
muon channel only. The inner error bar indicates the statistical 
uncertainty, the outer error 
bar the quadratic sum of the statistical and systematic uncertainties. This is
also true for the results from H1, FTPS and EMC Collaborations. The other 
three ZEUS measurements come from the combination of the electron and muon
results as described in the text. The inner error bars represent 
the statistical and decay 
channel specific errors added in quadrature, the outer ones the statistical, 
decay specific and common systematic errors added in quadrature. 
The lines correspond to the NLO prediction from
\protect\cite{kramer} assuming the GRV \protect\cite{grv} (continuous), 
MRSG \protect\cite{mrs2} (dashed) and CTEQ3M \protect\cite{cteq} (dotted-dashed) 
gluon distributions with $m_c=1.4$ GeV and $\Lambda_{QCD}=300$ MeV, 
the dotted curve was obtained with GRV, $m_c=1.55$ GeV and 
$\Lambda_{QCD}=215$ MeV . The curves are scaled up by a factor 
of 1.15 to take into 
account the contribution from $\psi' \rightarrow \mbox{J}/\psi X$.} 
\label{Fig:xsect}
\end{figure}

The electro- and photoproduction cross sections are summarized in Tab. 
\ref{Tab:xsect} for the two J/$\psi$ decay channels. The first 
error is statistical and the second comes 
from adding in quadrature all the systematic errors described in section 
\ref{s:sys}. 
The cross sections measured in the restricted $z$ ranges were 
extrapolated to 
$z = 0$ (using HERWIG) in order to be able to compare with other available 
data. The size of the extrapolation 
($\sim 10\%$ for the muon decay mode and $\sim 25\%$ for the electron decay
mode) 
and the associated systematic error are shown in Tab. \ref{Tab:xsect} and 
given in section \ref{s:sys}. The difference in the size of the 
extrapolation between muon and electron decay modes is due to the different 
$z$ ranges measured. 
Table \ref{Tab:xsectcomb} reports for the lowest $W$ bin the photoproduction 
cross section from the muon channel only and for the following three 
$W$ bins the combined muon and electron photoproduction 
cross sections. These results are given both in the range $0.4 < z < 0.9$ 
and in the extrapolated range $z < 0.9$. To obtain the combined results 
a weighted mean was calculated; the weights were obtained by
summing the statistical and decay channel specific errors in quadrature. The 
first error for the combined results in Tab. \ref{Tab:xsectcomb} is 
the error on the weighted mean, the second is given by the sum of the common
systematic errors added in quadrature. 
The photoproduction cross sections for $z < 0.9$ are shown in 
Fig. \ref{Fig:xsect} together with those found by the 
H1 collaboration \cite{H1} and a compilation of fixed target 
results \cite{FTPS,NA14,EMC2}. The ZEUS and H1 data are compatible. 
The cross section rises as $W$ increases. The curves in the plot 
correspond to the NLO
calculation \cite{kramer} computed with no cut on the $p_T$ of 
the \jpsi, $z < ~0.9$, with a charm mass ($m_c$) of 1.4 GeV, 
$ \Lambda_{QCD} = 300$ MeV and with 
renormalization and factorization scales of $\sqrt{2} m_c$. 
Dashed, continuous and dashed-dotted curves are obtained with  
different gluon distributions  
(MRSG \cite{mrs2}, GRV \cite{grv} and CTEQ3M \cite{cteq}, respectively), 
compatible with those extracted from $F_2$ 
measurements at HERA \cite{F2ZEUS,F2H1}. The theoretical 
predictions were multiplied by a factor 1.15  to take into account 
$\psi'$ production. The predictions are in qualitative 
agreement with the data, but the distinction among different parton densities 
is not possible because the NLO calculation is not well behaved in the limit 
$p_T \rightarrow 0$. Furthermore, there is a 
significant dependence on the 
values of the charm mass and $\Lambda_{QCD}$. The cross section 
varies as $1/m_c^3$ and $\alpha_s^2$ as illustrated by the dotted line which is
calculated with $m_c=1.55$ GeV and $\Lambda_{QCD}$ = 215 MeV and using the 
GRV gluon distribution.

\begin{figure}[b]
\epsfig{file=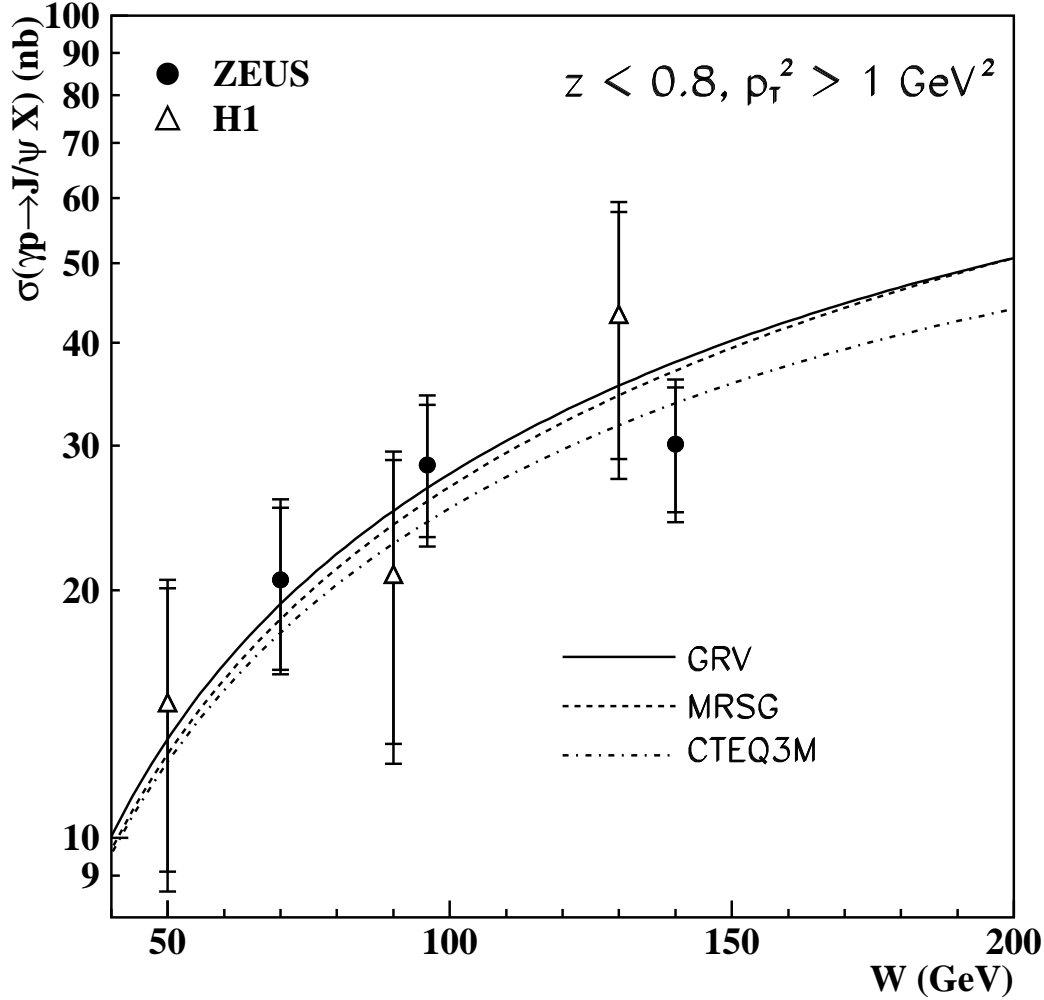,%
      width=15cm%
        }
\caption{The direct inelastic J/$\psi$ photoproduction cross section 
as a function of $W$ for $z<0.8$ and $p_T^2 > 1$ GeV$^2$.  
Data from ZEUS and H1 \protect\cite{H1} are shown. The ZEUS
results in the lowest two $W$ bins are obtained with the muon channel only. 
The inner error bars indicate the statistical uncertainties, the outer 
error bars the quadratic sum of the statistical and systematic uncertainties. 
This is also true for the results from the H1 Collaboration. 
The ZEUS measurement in the highest $W$ bin comes from the combination 
of the electron and muon results as described in the text. The inner error bar 
represents the statistical and decay 
channel specific errors added in quadrature, the outer one the statistical, 
decay specific and common systematic errors added in quadrature.
The lines correspond to the NLO prediction from
\protect\cite{kramer} assuming the GRV \protect\cite{grv} (continuous), 
MRSG \protect\cite{mrs2} (dashed) and CTEQ3M \protect\cite{cteq} (dotted-dashed) 
gluon distributions with $m_c=1.4$ GeV and $\Lambda_{QCD}=300$ MeV.
The curves are scaled up by a factor of 1.15 to take into 
account the contribution from $\psi' \rightarrow \mbox{J}/\psi X$.} 
\label{Fig:xsect08}
\end{figure}

A more quantitative comparison between data and theory can be made in the
restricted kinematic range $p_T^2 > ~1$ GeV$^2$,  where the 
calculation is much more reliable. The NLO 
computation now allows an absolute comparison between data and 
models. 
The cross sections for this kinematic range are summarized 
in Tab. \ref{Tab:xsect} for the two J/$\psi$ decay channels and in 
Tab. \ref{Tab:xsectcomb} for the combined result, with the 
additional requirement $z<0.8$ in order to compare with \cite{H1}. 
The measured cross sections for $z < 0.8$ and 
$p_T^2 > ~1$ GeV$^2$ are displayed in Figure \ref{Fig:xsect08}. 
The curves represent the NLO calculation 
using the different gluon distributions cited above. 
Data and theory are in good agreement  
using $m_c=1.4$ GeV, $\Lambda_{QCD}=300$ MeV and $\sqrt{2} m_c$ as 
renormalization factor.   
While for $z<0.9$ and all $p_T^2$ the predictions are significantly 
different for different parametrizations of the gluon density, 
this is not so 
in the more restricted domain $z< ~0.8$ and $p_T^2 > ~1$ GeV$^2$. 
This is a consequence of the $p_T^2$ cut: 
the gluon distribution is probed at larger 
$x_g$ values, where the differences between the various gluon densities 
are smaller; here $x_g$ is the proton energy fraction carried by the 
incoming gluon. In the present analysis, with 
$z<0.9$ we explore the range $4\cdot 10^{-4} \ltap ~x_g~ \ltap 10^{-2}$, 
and the range $10^{-3} \ltap ~x_g~ \ltap 10^{-2}$ in the restricted interval.

\subsection{ Transverse Momentum Distribution}
\label{s:pt2trans}

\begin{figure}[b]
\epsfig{file=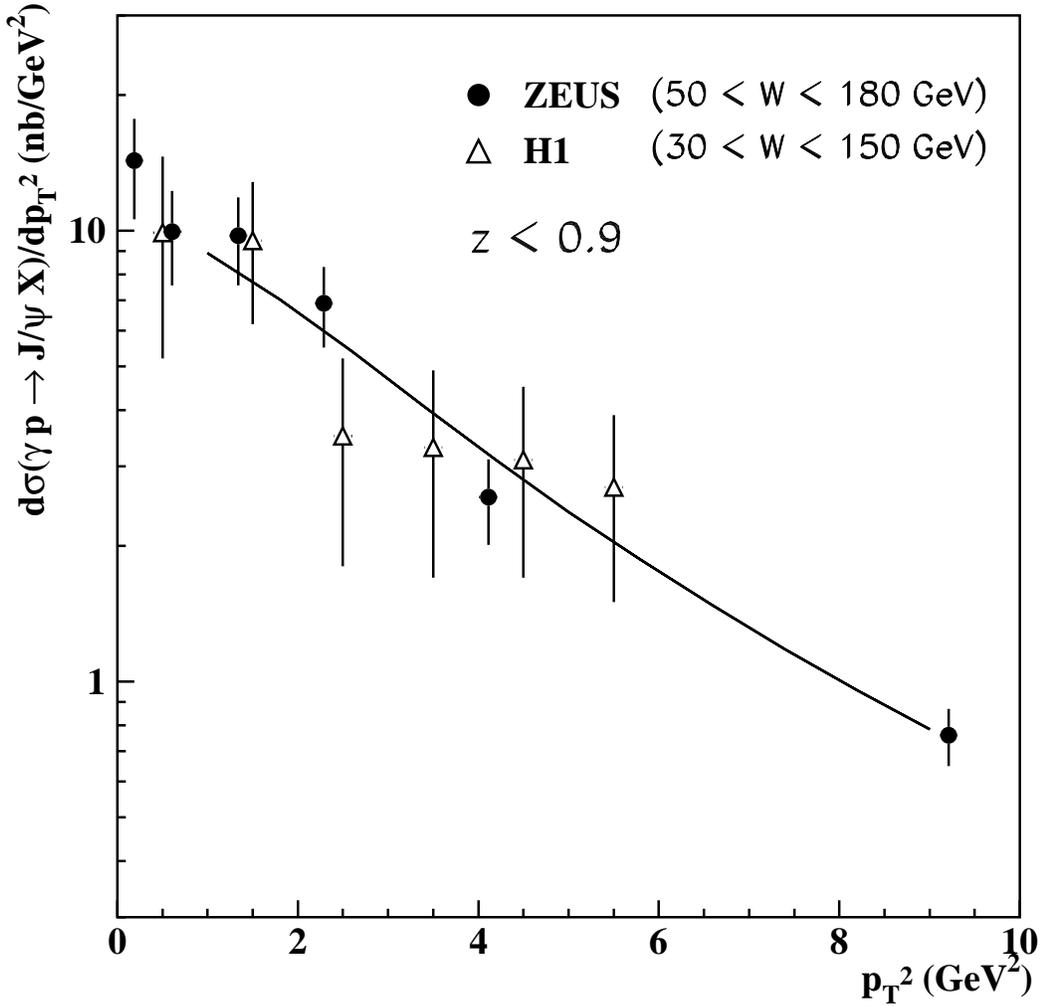,%
      width=15cm%
        }
\caption{Differential cross section $d\sigma/dp_T^2$ for the inelastic 
J/$\psi \rightarrow \mu^{+}\mu^{-}$ sample with $50 < W < 180$ GeV and 
$z < 0.9$. Data from ZEUS and H1 \protect\cite{H1} are shown. 
The error bars indicate 
the quadratic sum of the statistical and systematic uncertainties.
The NLO computation \protect\cite{kramer} with the GRV \protect\cite{grv} 
structure function, $m_c=1.4$ GeV and $\Lambda_{QCD}=300$ MeV 
is shown as the solid line. The theoretical curve is drawn 
only for $p_T^2 > 1$ GeV$^2$ because in the low $p_T$ region the calculation is not 
reliable. 
In the theoretical curve the 15\% contribution of the $\psi'$ has not been 
included.}
\label{Fig:pt2diff}
\end{figure}

Figure \ref{Fig:pt2diff} shows the differential cross section 
d$\sigma$/d$p_T^2$ for $z < 0.9$ and $50 < W <180$ GeV 
using only the muon sample. The background contributions listed in section 
\ref{s:back} were subtracted
bin by bin. The curve shows the  NLO prediction obtained with 
$m_c=1.4$ GeV, $\Lambda_{QCD}=300$ MeV, $\sqrt{2} m_c$ as 
scale and GRV for the gluon distribution\footnote{The NLO $p_T^2$ and $z$ 
distributions for the $\psi'$ have large theoretical uncertainties and cannot 
be accounted for simply by a scale factor.}.
For $p_T^2 > 1$ GeV$^2$ data and theoretical calculation are in 
good agreement. A fit of the function
\begin{equation}
    \frac{d \sigma}{dp_{T}^2} = A e^{-b p_{T}^2}
\label{Eq:dsdpt2}
\end{equation}
to our data was performed in the range $1 < p_T^2 < 9$ GeV$^2$ giving
\begin{equation}
b = 0.32 \pm 0.03~ \mbox{(stat)} \pm 0.01 \mbox{(syst)}~ \mbox{GeV}^{-2}.
\end{equation}
A similar fit to the NLO calculation \cite{kramer} yields a slope of 
$b=0.3$ GeV$^{-2}$ above $p_T^2 > 1$ GeV$^2$. The systematic error contains 
contributions from all the classes of systematic errors discussed in 
section \ref{s:sys} and also from the change in the $p_T^2$ fitting 
interval.

\subsection{  Distribution of $z$}
\label{s:zdist}

\begin{figure}[b]
\epsfig{file=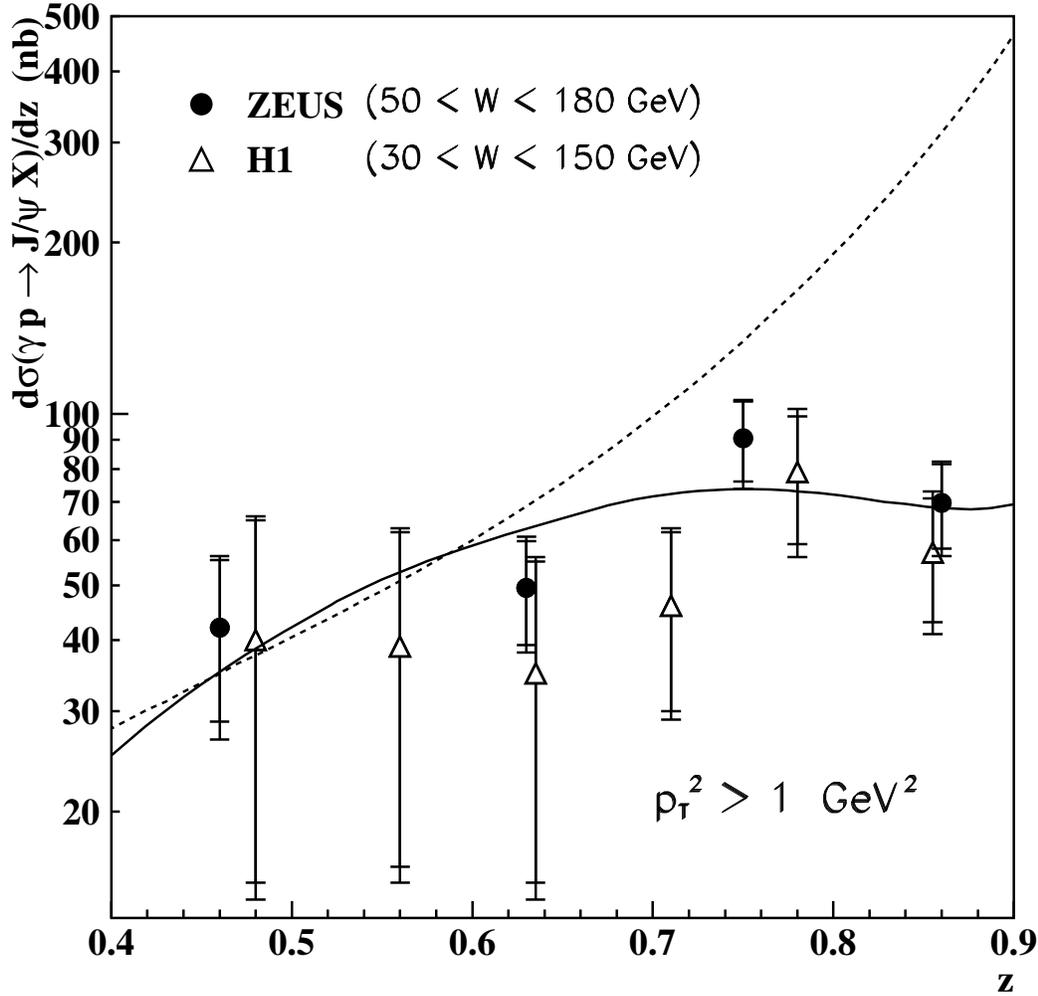,%
      width=15cm%
        }
\caption{Differential cross section $d\sigma/dz$ for the inelastic 
J/$\psi \rightarrow \mu^+ \mu^-$ sample with $50 < W < 180$ GeV and 
$p_T^2 > 1$ GeV$^2$. 
Data from ZEUS and H1 \protect\cite{H1} are shown.
The inner error bars indicate the statistical uncertainties, the outer error 
bars the quadratic sum of the statistical and systematic uncertainties.
The NLO computation \protect\cite{kramer} with the GRV \protect \cite{grv} 
structure function, $m_c=1.4$ GeV and $\Lambda_{QCD}=300$ MeV   
is shown as a solid line. The dashed line is given by the 
sum of the colour-singlet and the colour-octet leading order 
calculations \protect\cite{cac}. In the theoretical curves 
the 15\% contribution of the $\psi'$ has not been included.}
\label{Fig:zdiff}
\end{figure}

Figure \ref{Fig:zdiff} shows the differential cross section d$\sigma$/d$z$ 
for $p_T^2 > 1$ GeV$^2$ and $50 < W < 180$ GeV as obtained using only the 
muon sample. 
It is compared to the NLO calculation \cite{kramer} discussed  
in section \ref{s:section} and with the 
parameters used in \ref{s:pt2trans}. 
Agreement in shape and normalization
is found within the errors. Our data are in good agreement with the 
result of the H1 collaboration \cite{H1}.  

Recently there has been theoretical activity 
attempting to solve the discrepancy between the 
J$/\psi$ production cross section measurements in hadronic reactions and the 
colour-singlet model by invoking additional octet 
contributions \cite{octet2}.
A specific leading order calculation of \jpsi photoproduction at HERA has 
been carried out using values of the nonperturbative colour-octet terms 
determined from a fit \cite{cac} to the CDF data \cite{CDF}. 
These calculations 
predict a cross section for HERA rising with $z$, which is not seen in the data. 
This is illustrated in Fig.\ref{Fig:zdiff} where the dashed line shows a sum 
of the colour-singlet and colour-octet contributions both calculated at 
leading order.
  
%The same contributions have been calculated in 
%leading order for J$/\psi$ photoproduction at HERA \cite{cac} using 
%the values of the 
%nonperturbative colour-octet terms determined from fits to the 
%CDF data \cite{CDF}. 
%These calculations predict a large inelastic contribution for $z >0.8$. 
%In Fig. \ref{Fig:zdiff} the dashed line is given by the sum of the
%colour-singlet and of the colour-octet leading order calculations.  
%The strong increase of the cross section with $z$ is not supported by 
%our data. 

\section{\bf Conclusion}
\label{Sect:Conclusion}

We have measured inelastic \jpsi photoproduction in the range 
$50 < W < 180$ GeV and $0.4 < z < 0.9$. The cross section rises 
with $W$. In this $z$ interval the photon-gluon fusion process is 
expected to dominate. 
A NLO calculation for the photon-gluon fusion process agrees with the data both 
for the integrated cross section and 
the differential distributions over $z$ and $p_T^2$ in the kinematic 
range $z <0.8$ and $p_T^2 > 1 $ GeV$^2$, using gluon distribution 
parametrizations compatible with those determined from the $F_2$ 
measurements performed at HERA. 
The predictions of a specific leading order colour-octet model, as 
formulated to fit the CDF data on J/$\psi$ hadroproduction, are not 
consistent with the data.

\vspace{.5 cm}
\noindent {\bf\Large Acknowledgements}
 
We thank the DESY Directorate for their strong support and encouragement.
The remarkable achievements of the HERA machine group were essential for the
successful completion of this work and are gratefully acknowledged.
We are also grateful to M. Cacciari, M. Kr\"amer, and P. A. Zerwas for 
providing the theoretical curves and for many useful discussions.

%
%       Reference
%

\newpage
\begin{table}
\begin{center}
\begin{tabular}{|l|c|c|c|c|} \hline
\multicolumn{5}{|c|}{  J/$\psi \rightarrow \mu^{+}\mu^{-}$  } \\ \hline\hline
$W$ range (GeV)    & 50-90    & 90-120   & 120-150   & 150-180 \\ \hline
$0.9< z < 1$  & 14$\pm$4 & 29$\pm$6 & 27$\pm$6  & 12$\pm$5 \\ \hline
$0.4< z < 0.9$ & 67$\pm$9 & 53$\pm$8 & 35$\pm$7  & 26$\pm$7 \\ \hline
$z<0.4$ & \multicolumn{4}{|c|}{19$\pm6$} \\ \hline\hline  
\multicolumn{5}{|c|}{ J/$\psi \rightarrow e^{+}e^{-}$  } \\ \hline\hline
$W$ range (GeV)    & 50-90    & 90-120   & 120-150   & 150-180 \\ \hline 
$0.9< z < 1$  &       & \multicolumn{3}{|c|}{22 $\pm$ 6} \\ \hline
$0.5< z < 0.9$ &       & 20$\pm$6 & 33$\pm$7  & 7$\pm$3 \\ \hline
\end{tabular}
\caption{Number of events in the various $z$ and $W$ ranges for 
J$/\psi \rightarrow \mu^{+}\mu^{-}$ and $e^{+}e^{-}$.}
\label{Tab:events}
\end{center}
\end{table}

\begin{table}
\begin{center}
\begin{tabular}{|c|c|c|c|c|c|c|} \hline
%\begin{tabular}{|p{1.8cm}|p{1.5cm}|p{0.8cm}|p{3.5cm}|p{2.6cm}|p{1.2cm}|p{2.6cm}|}
\hline
\multicolumn{7}{|c|}{ J/$\psi \rightarrow \mu^{+}\mu^{-}$, $z < 0.9$} 
\\ \hline\hline
{\small $W$ } & {\small $<W>$ } & {\small $\cal{A}$ } & 
{\small $\sigma_{ep}$(0.4$<z<$0.9)} &
{\small $\sigma_{ep}$($z<$0.9)} & {\small $\Phi_T$ } &  
{\small $\sigma_{\gamma p}$($z<$0.9) }   
\\ 
{\small(GeV)}  & {\small(GeV)} &  & {\small(nb)} & {\small(nb)} &  & 
{\small(nb)}  \\ \hline
50-90   &  73 & 18$\%$ & 1.74$\pm0.23^{+0.19}_{-0.14}$ & 1.92$\pm0.25^{+0.22}_{-0.17}
$ & 0.0555 & 34.6$\pm4.5^{+4.0}_{-3.1}$ \\ \hline
90-120  & 105 & 24$\%$ & 1.10$\pm0.17^{+0.10}_{-0.07}$ & 1.23$\pm0.19^{+0.12}_{-0.09}
$ & 0.0232 & 53.0$\pm8.2^{+5.2}_{-4.0}$ \\ \hline
120-150 & 134 & 24$\%$ & 0.75$\pm0.14^{+0.08}_{-0.04}$ & 0.84$\pm0.16^{+0.09}_{-0.06}
$ & 0.0157 & 53.5$\pm10.2^{+5.7}_{-3.8}$ \\ \hline
150-180 & 162 & 16$\%$ & 0.81$\pm0.20^{+0.08}_{-0.07}$ & 0.90$\pm0.22^{+0.10}_{-0.09}
$ & 0.0110 & 81.8$\pm20.0^{+9.1}_{-8.2}$ \\ \hline\hline    
\multicolumn{7}{|c|}{ J/$\psi \rightarrow e^{+}e^{-}$, $z < 0.9$} 
\\ \hline\hline 
{\small $W$} & {\small $<W>$ } & {\small $\cal{A}$ } & 
{\small $\sigma_{ep}$(0.5$<z<$0.9)} &
{\small $\sigma_{ep}$($z<$0.9)} & {\small $\Phi_T$ } &  
{\small $\sigma_{\gamma p}$($z<$0.9) }
\\ 
{\small(GeV)}  & {\small(GeV)} &  & {\small(nb)} & {\small(nb)} &  & 
{\small(nb)}  \\ \hline
90-120  & 107 & 16$\%$ & 0.63$\pm0.20^{+0.10}_{-0.06}$ & 0.79$\pm0.25^{+0.14}_{-0.08}
$ & 0.0232 & 34.1$\pm10.8^{+6.0}_{-3.4}$ \\ \hline
120-150 & 136 & 20$\%$ & 0.85$\pm0.18^{+0.10}_{-0.09}$ & 1.05$\pm0.22^{+0.13}_{-0.13}
$ & 0.0157 & 66.9$\pm14.0^{+8.3}_{-8.3}$ \\ \hline
150-180 & 166 &  7$\%$ & 0.55$\pm0.25^{+0.07}_{-0.09}$ & 0.68$\pm0.31^{+0.09}_{-0.11}
$ & 0.0110 & 61.8$\pm28.2^{+8.2}_{-10.0}$ \\ \hline\hline
\multicolumn{7}{|c|}{ J/$\psi \rightarrow \mu^{+}\mu^{-}$, $z < 0.8$, 
$p_{T}^2 > 1~$GeV$^2$} 
\\ \hline\hline
{\small $W$} & {\small $<W>$ } & {\small $\cal{A}$ } & 
{\small $\sigma_{ep}$(0.4$<z<$0.8)} &
{\small $\sigma_{ep}$($z<$0.8)} & {\small $\Phi_T$ } &  
{\small $\sigma_{\gamma p}$($z<$0.8) }   
\\ 
{\small(GeV)}  & {\small(GeV)} &  & {\small(nb)} & {\small(nb)} &  & 
{\small(nb)}  \\ \hline
50-80   &  70 & 16$\%$ & 0.80$\pm0.18^{+0.09}_{-0.05}$ & 0.93$\pm0.21^{+0.11}_{-0.06}
$ & 0.0452 & 20.6$\pm4.6^{+2.5}_{-1.4}$ \\ \hline      
80-110  &  96 & 22$\%$ & 0.65$\pm0.12^{+0.07}_{-0.05}$ & 0.76$\pm0.14^{+0.09}_{-0.07}
$ & 0.0268 & 28.4$\pm5.2^{+3.2}_{-2.5}$  \\ \hline
110-180 & 138 & 18$\%$ & 0.89$\pm0.19^{+0.09}_{-0.08}$ & 1.04$\pm0.22^{+0.12}_{-0.10}
$ & 0.0334 & 31.1$\pm6.6^{+3.5}_{-3.1}$  \\ \hline\hline
\multicolumn{7}{|c|}{ J/$\psi \rightarrow e^{+}e^{-}$, $z < 0.8$, $p_{T}^2 > 1~$GeV$^2$}
\\ \hline\hline 
{\small $W$} & {\small $<W>$ } & {\small $\cal{A}$ } & 
{\small $\sigma_{ep}$(0.5$<z<$0.8)} &
{\small $\sigma_{ep}$($z<$0.8)} & {\small $\Phi_T$ } &  
{\small $\sigma_{\gamma p}$($z<$0.8) }  
\\ 
{\small(GeV)}  & {\small(GeV)} &  & {\small(nb)} & {\small(nb)} &  & 
{\small(nb)}  \\ \hline
110-180 & 142 & 15$\%$ & 0.70$\pm0.18^{+0.08}_{-0.09}$ & 0.96$\pm0.25^{+0.12}_{-0.13}
$ & 0.0334 & 28.7$\pm7.5^{+3.6}_{-3.9}$  \\ \hline
\end{tabular}
\caption{Inelastic J$/\psi$ photoproduction cross sections for the muon 
and electron decay modes in the two regions $z <0.9$ and $z<0.8$, $p_T^2 > 1$ GeV$^2$. 
From left to right we give the $W$ range, the 
$W$ mean value ($<W>$), the acceptance ($\cal{A}$), the measured $ep$ cross sections,
the $ep$ cross sections extrapolated to $z=0$, the flux
factor ($\Phi_T$) and finally the $\gamma p$ cross sections.}
\label{Tab:xsect}
\end{center}
\end{table}

\newpage
\begin{table}
\begin{center}
\begin{tabular}{|c|c|c|c|} \hline
$W$ & $<W>$ & $\sigma_{\gamma p}$ ($0.4< z <0.9$) & 
$\sigma_{\gamma p}$ ($z <0.9$) \\ 
{\small(GeV)} & {\small(GeV)} &{\small(nb)} & {\small(nb)}  \\ \hline\hline
50-90   &  73 & 31.4$\pm4.1^{+3.4}_{-2.5} $ & 34.6$\pm4.5^{+4.0}_{-3.1}$ \\ \hline
90-120  & 106 & 41.4$\pm6.0^{+3.0}_{-1.3} $ & 46.1$\pm7.0^{+3.6}_{-2.2}$ \\ \hline
120-150 & 135 & 51.0$\pm7.6^{+4.5}_{-1.9} $ & 57.8$\pm8.7^{+5.6}_{-3.0}$ \\ \hline
150-180 & 162 & 68.2$\pm14.5^{+6.4}_{-5.5}$ & 75.3$\pm16.6^{+7.4}_{-6.9}$ \\ \hline\hline 
$W$ & $<W>$ & $\sigma_{\gamma p}$ ($0.4 < z <0.8, p_T^2 > 1$ GeV$^2$) & 
$\sigma_{\gamma p}$ ($z <0.8, p_T^2 > 1$ GeV$^2$) \\ 
{\small(GeV)} & {\small(GeV)} &{\small(nb)} & {\small(nb)}  \\ \hline\hline
50-80   &  70 & 17.7$\pm4.0^{+2.0}_{-1.1}$ & 20.6$\pm4.6^{+2.5}_{-1.4}$ \\ \hline
80-110  &  96 & 24.3$\pm4.5^{+2.6}_{-1.9}$ & 28.4$\pm5.2^{+3.2}_{-2.5}$ \\ \hline
110-180 & 140 & 25.7$\pm4.2^{+2.4}_{-2.1}$ & 30.1$\pm5.2^{+3.0}_{-2.7}$ \\ \hline
\end{tabular}
\caption{Cross sections $\sigma_{\gamma p \rightarrow J/\psi X}$ 
for the phase space regions: $0.4 < z < 0.9$ (top-left), $z <0.9$ (top-right), 
$0.4 < z< 0.8$ and $p_T^2 > 1$ GeV$^2$ (bottom-left), 
$z< 0.8$ and $p_T^2 > 1$ GeV$^2$ (bottom-right). 
The cross sections for $0.4 < z < 0.9$ and for $z < 0.9$ in the $W$ range from 
50 to 90 GeV come from the muon channel only. 
The first error is statistical, the second one
comes from all the systematic errors added in quadrature. The other three 
measurements come from the combination of the electron and muon 
results as described in the text. The first error contains the contribution 
from statistical and decay channel specific errors while the second 
contains all sources of common systematic errors. 
In the regions $0.4 < z < 0.8$,  
$p_T^2 > 1$ GeV$^2$ and $z < 0.8$, $p_T^2 > 1$ GeV$^2$ for $W$ 
in the range from 50 to 110 GeV only the muon
channel data are used while for the highest bin electron and muon 
results were combined as explained in the text.}
\label{Tab:xsectcomb}
\end{center}
\end{table}

\newpage

\newpage

\newpage

\newpage

\newpage

\newpage

\newpage

\end{document}